\documentclass[twocolumn,amsmath,amssymb,floatfix,superscriptaddress,prl]{revtex4-2}
\usepackage{physics}
\usepackage[colorlinks=true,linkcolor=blue,citecolor=blue]{hyperref}
\usepackage{bm,graphicx,epsfig}
\usepackage[usenames]{color}
\usepackage[english]{babel}
\usepackage{amsfonts,stmaryrd,amssymb}
\usepackage{enumerate} 
\usepackage[ruled]{algorithm2e} 
\usepackage[framemethod=tikz]{mdframed} 
\usepackage{amsthm,amsmath,amssymb,amsfonts,graphicx,verbatim,xcolor,bm} 
\usepackage{subcaption}
\usepackage[justification=RaggedRight]{caption}
\usepackage[normalem]{ulem}

\usepackage{dsfont}

\usepackage{csquotes}
\MakeOuterQuote{"}

\allowdisplaybreaks

\begin{document}

	\newcommand{\Kelvin}{\text{K}}
	\newcommand{\nm}{\text{nm}}
	\newcommand{\meV}{\text{meV}}
	\newcommand{\eV}{\text{eV}}
	\newcommand{\bR}{\mathbf{R}}
	\newcommand{\bX}{\mathbf{X}}
	\newcommand{\bM}{\mathbf{M}}
	\newcommand{\bQ}{\mathbf{Q}}
	\newcommand{\bG}{\mathbf{G}}
	\newcommand{\bK}{\mathbf{K}}
	\newcommand{\bq}{\mathbf{q}}
	\newcommand{\bk}{\mathbf{k}}
	\newcommand{\bp}{\mathbf{p}}
	\newcommand{\bL}{\mathbf{L}}
	\newcommand{\bl}{\mathbf{l}}
	\newcommand{\bx}{\mathbf{x}}
	\newcommand{\by}{\mathbf{y}}
	\newcommand{\bz}{\mathbf{z}}
	\newcommand{\br}{\mathbf{r}}
	\newcommand{\eqn}[1]{(\ref{#1})}
	\newcommand{\dg}[1]{{\color{orange}[\textbf{DG: }\textit{{#1}}]}}
	\newcommand{\ajm}[1]{{\color{purple}[\textbf{AJM: }\textit{{#1}}]}}
	\newcommand{\mr}{moir\'e~}
	\newcommand{\Mr}{Moir\'e~}
	\newcommand{\mC}{\mathcal{C}}
	\newcommand{\mT}{\mathcal{T}}
	\newcommand{\mM}{\mathcal{M}}
	\newcommand{\be}{\begin{equation}}
		\newcommand{\ee}{\end{equation}}
	\newenvironment{eqs}
	{\begin{equation} \begin{aligned}}
			{\end{aligned} \end{equation} }
	\newcommand{\bal}{\begin{eqs}}
		\newcommand{\eal}{\end{eqs}}
	\newcommand{\mA}{\mathcal{A}}
	\newcommand{\mB}{\mathcal{B}}
	\newcommand{\mF}{\mathcal{F}}
	\newcommand{\CommentDG}[1]{\textbf{\color{orange} [DG: #1]}}
	\newcommand{\nw}[1]{\textbf{\color{blue} [NW: #1]}}
	\newcommand{\gs}[1]{\textbf{\color{red} [GS: #1]}}
	\newcommand{\todo}[1]{\textbf{\color{red} [To do: #1]}}

	\title{Edge zeros and boundary spinons in topological Mott insulators}

	\author{Niklas Wagner}
	\thanks{These authors contributed equally to this work.}
		\affiliation{Institut f\"ur Theoretische Physik und Astrophysik and
W\"urzburg-Dresden Cluster of Excellence ct.qmat, Universit\"at W\"urzburg, 97074 W\"urzburg, Germany}
	
	\author{Daniele Guerci}
	\thanks{These authors contributed equally to this work.}
	\affiliation{Center for Computational Quantum Physics, Flatiron Institute, New York, NY 10010, USA }
	
	\author{Andrew J. Millis}
	\affiliation{Department of Physics, Columbia University, New York, NY 10027, USA} 
	\affiliation{Center for Computational Quantum Physics, Flatiron Institute, New York, NY 10010, USA }
	
	\author{Giorgio Sangiovanni}
	\email{sangiovanni@physik.uni-wuerzburg.de}
	\affiliation{Institut f\"ur Theoretische Physik und Astrophysik and
W\"urzburg-Dresden Cluster of Excellence ct.qmat, Universit\"at W\"urzburg, 97074 W\"urzburg, Germany}
	
	\date{\today}

	\begin{abstract}
		
		
		We introduce a real-space slave rotor theory  of the physics of topological Mott insulators, using the Kane-Mele-Hubbard model as an example, and use it to show that a topological gap in the Green function zeros corresponds to a gap in the bulk spinon spectrum and that a zero edge mode corresponds to a spinon edge mode. We then consider  an interface between a topological Mott insulator and a conventional topological insulator showing how  the spinon edge mode of the topological Mott insulator combines with the spin part of the conventional electron topological edge state leaving a non-Fermi liquid edge mode described by a gapless propagating holon and gapped spinon state.  Our work demonstrates the physical meaning of Green function zeros and shows that  interfaces between  conventional and Mott topological insulators are a rich source of new physics.

	\end{abstract}

	\maketitle

	\noindent
	{\it Introduction --}  A conventional metal is characterized by gapless low energy electronic quasiparticles with charge $e$ and spin $1/2$ and an electronic Green function with poles that define the Fermi surface. A conventional topological insulator is characterized by a gap to fermionic (or other) excitations and a gapless fermionic edge state. A strongly correlated Mott insulator  may exhibit fractionalized quasiparticles (spinons) with charge zero but spin $1/2$ and also an electronic Green function characterized with a surface of zeros that defines a ``Luttinger surface'' that generalizes the concept of the Fermi surface to the Mott state ~\cite{oshikawa_topological_2000,senthil_fractionalized_2003,paramekanti_extending_2004,rosch_breakdown_2007,sakai_evolution_2009,sakai_doped_2010,volovik_topology_2012,seki_topological_2017,heath_necessary_2020,wen_low-energy_2021,else_non-fermi_2021,skolimowski_luttingers_2022}. 

 Recently it has been shown that Green function zeros play a crucial role also in strongly interacting topological Mott insulators
 ~\cite{fabrizio_landau-fermi_2020,fabrizio_emergent_2022,wagner_mott_2023,setty_electronic_2023,blason_unified_2023,setty_symmetry_2023,setty_topological_2023}.  In particular, it is possible to define an interacting topological invariant~\cite{ishikawa_magnetic_1986,ishikawa_microscopic_1987,gurarie_single-particle_2011,essin_bulk-boundary_2011,wang_simplified_2012,slagle_2015},  in terms of the fully interacting single particle Green function in a manner that  treats  poles and zeros of the Green function on equal footing and, moreover, extends the bulk-boundary correspondence  to the strongly interacting regime simply replacing edge poles with zeros~\cite{essin_bulk-boundary_2011,manmana_topological_2012,wagner_mott_2023}.
 Recently, some of us and others conjectured~\cite{wagner_mott_2023}, based on  analysis of one dimensional models, that at the interface between a Mott and conventional topological insulator the gapless edge zero band could annihilate the edge state of the conventional topological insulator. However, the physical meaning of the edge zeros has remained unclear  because the zeros do not directly correspond to any physical excitation and   the topological invariant is not directly connected to the Hall conductance of the strongly interacting system~\cite{blason_unified_2023,gavensky_connecting_2023,setty_electronic_2023,zhao_failure_2023}.
 
	
	
	 
	\begin{figure}[h!]
		\includegraphics[width=0.8\linewidth]{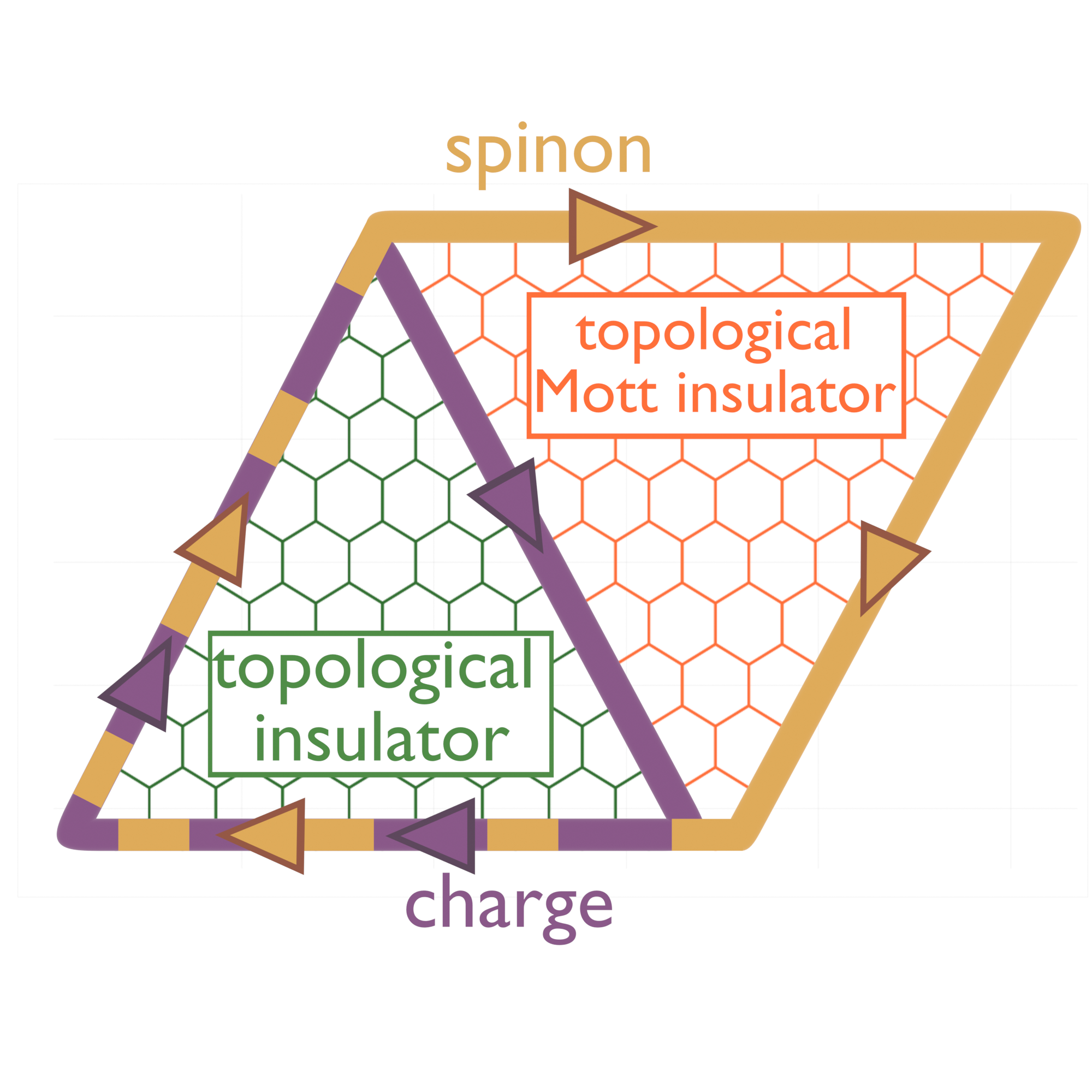}
		\caption{Illustration of the spin and charge separation at an interface between a topological Mott insulator (TMI) and weakly-interacting topological insulator (TI). The edge state of the TI consists of a spin and charge contribution whereas the TMI exhibits only a spinon edge state. Along the interface the spinon channels compensate leaving only a charge channel. Note that only one spin channel is shown.}
		\label{fig:spin-charge-separation}
	\end{figure}
	

	In this work we use a real-space auxiliary particle method to connect the properties of Green function zeros of a Mott insulator to the properties of spinon excitations, in particular demonstrating a one-to-one correspondence between bulk gaps in the Green function zeros and in the spinon spectrum and between gapless boundary zeros and gapless boundary spinons.  Moreover we give a new perspective on the physics emerging at the interface~\cite{aasen_electrical_2020} between a topological Mott insulator \cite{raghu_topological_2008,maciejko_fractionalized_2015} and a conventional topological insulator, showing (see Fig.~\ref{fig:spin-charge-separation}) how the gapless spinon \cite{pesin_mott_2010} (Green function zero) of the topological Mott insulator cancels the spinon portion of the conventional topological edge state, leaving a spinless holon excitation on the conventional topological insulator side. 
	

	{\it Model and Methods --} We consider the interacting Kane-Mele Hamiltonian $H=H_t+H_{\rm SOC}+H_{int}$ \cite{kane_z_2005}, which consists of electrons moving on sites of a lattice with two sublattices per unit cell and subject to a local on-site interaction. This model exhibits both metallic, conventional insulator and Mott insulating phases, which are  topological for $\lambda\neq 0$.   This model is of current interest as a model of transition metal dichalcogenides moir\'e semiconductors~\cite{zhao_realization_2022,mak_semiconductor_2022}. The Hamiltonian terms are: 
 \begin{equation}
    \begin{split}		
  H_t&=-t\sum_{\langle\br\in A,\br'\in B\rangle}f^{A\dagger}_{\br}f^B_{\br'}+h.c. ,\\
		H_{\rm SOC}&=\lambda\sum_{s=A,B} \sum_{\langle \br,\br'\rangle\in s} f^{s\dagger}_{\br} e^{i\frac{\pi}{2} \nu_{\br,\br'}\sigma^z}f^s_{\br'},
    \end{split}	
\end{equation}
	where $f^s_{\br}=[f^s_{\br\uparrow},f^s_{\br\downarrow}]^T$ is a two-dimensional spinor, ${s=A,B}$ refers to the two inequivalent sublattices, $\sigma^z$ is the Pauli matrix in the spin space and $\nu_{\br,\br'}=+1(-1)$ if the path connecting $\br$ to $\br'$ winds counterclockwise (clockwise). In this study we do not consider the effect of a staggered potential distinguishing the two sublattices.
 The Coulomb interaction is given by the on-site Hubbard term: 
	\begin{equation}
		H_{\rm int}=\frac{U}{2}\sum_{s=A,B}\sum_{\br\in s}\left(\sum_\sigma n^s_{\br\sigma}-1\right)^2,
	\end{equation}
	where $n_{\br\sigma}$ is the electron number at site $\br$ with spin $\sigma$. At $\lambda=0$ the model has two important symmetries: the discrete rotation $C_{3z}$ and the product of spinless time reversal and inversion $C_{2z}\mathcal{T}$ which imply a gapless (Dirac) spectrum for the weakly correlated half filled metallic phase and a gapless (Dirac) spinon spectrum in the Mott insulating phase. $\lambda\neq 0$ breaks $C_{2z}\mathcal T$, gapping the bulk electron (weakly correlated) or spinon (strongly correlated) spectra in the half filled case ~\cite{haldane_model_1988,kane_z_2005}.

	We are interested in studying the model for large interaction $U\gg t$ in the paramagnetic Mott insulating regime considered also in Ref.~\cite{rachel_topological_2010,rachel_interacting_2018,fernandez_lopez_bad_2022}. 
	We employ the slave-rotor technique which consists of introducing the O(2) rotor~\cite{florens_quantum_2002,florens_coherence_2003,florens_slave-rotor_2004}: 
	\begin{equation}
		\label{slave_rotor_mapping}
		f^s_{\br\sigma}=e^{i\theta^s_{\br}}\psi^s_{\br\sigma},
	\end{equation}
	where the phase degree of freedom $\theta^{s}_\br$ is conjugate to the local charge and $\psi^s_{\br\sigma}$ is an auxiliary fermion (spinon). The enlarged Hilbert space of rotors and spinons is constrained to $L^s_\br-\sum_\sigma \psi^{s\dagger}_{\br\sigma}\psi^s_{\br\sigma}+1=0$. We treat the theory in a saddle point approximation justified in a large spin degeneracy limit. Technical details of the calculations are given in the Supplementary Material ~\cite{Suppmat}. 
	
Although more accurate methods exist, this approach captures non perturbative phenomena such as the Mott transition~\cite{florens_slave-rotor_2004,lee_u1_2005} and offers a transparent physical interpretation of the correlated state. Moreover, the method is simple enough to allow simulations with different boundary conditions, that is crucial for establishing the connection between bulk topological invariants and edge modes of topological Mott insulators. 
	Within the self-consistent gaussian approximation, formally derived in the large-$M$ limit~\cite{florens_slave-rotor_2004}, the slave rotor method gives a non-trivial description of the Mott regime, including dispersive Hubbard bands, spinon excitations with hopping scaling as $\sim t^2/U$ and preserving many qualitative properties of the electronic spectral function~\cite{florens_quantum_2002,florens_coherence_2003,florens_slave-rotor_2004,lee_u1_2005}. This both allows the study of systems with an unprecedented level of reliability and enables studies of the finite-size geometries needed for a detailed analysis of the bulk-edge correspondence in the strongly correlated Mott regime.

	
	\noindent
	{\it Results -- } 
	We focus on the electron Green function $G_\sigma(\bk,\omega+i0^+)$ which in the slave rotor representation is a composite correlation function involving the spinon ($\psi^s$) and rotor ($\theta$) fields as $G^{ss'}_{\sigma}(\br-\br' ,\tau)=-\langle T_\tau\left(e^{i(\theta^{s}_{\br}(\tau)-\theta^{s'}_{\br'})}\psi^s_{\br\sigma}(\tau) \psi^{s'\dagger}_{\br'\sigma}\right)\rangle$. In the saddle-point approximation the Green function takes a simple form with an analytical expression detailed in the Supplementary Material~\cite{Suppmat} (see also Ref.~\cite{he_magnetic_2022,he_electronic_2023} for similar derivations): 
		\begin{equation}
			\begin{split}
				\label{green_physical_komega}
				G^{ss'}_\sigma(\bk,i\omega)&=Z^{ss'} G^{ss'}_{\sigma,\psi}(\bk ,i\omega)\\
				+ \frac{T}{N}&\sum_{\bq,i\nu} G^{ss'}_{\sigma,\psi}(\bk-\bq,i\omega-i\nu) G^{ss'}_X(\bq,i\nu),
			\end{split}
		\end{equation}
		where $\omega=i\pi(2n+1)T$, $\nu=i\pi 2n T$ with $T$ temperature and $G_\psi$ and $G_X$ are the Green function of the spinon and the rotor, respectively.  Finally, $Z^{ss'}$ is the quasiparticle weight which vanishes in the Mott insulator.

\begin{figure}[h]
		\centering
		\includegraphics[width=1\linewidth]{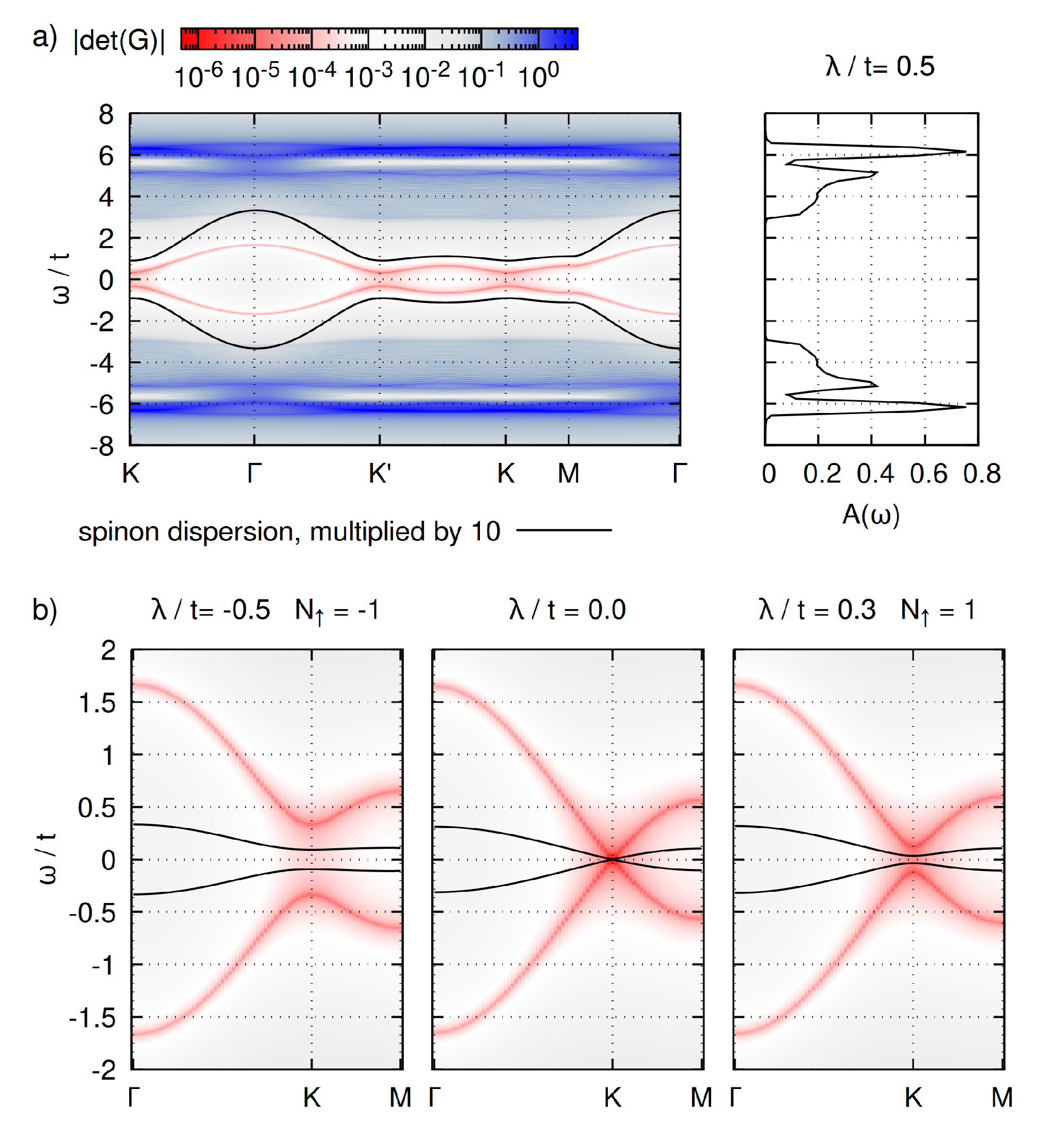}
		\caption{a) Left panel: color plot of the determinant of the slave-rotor Green function for $\lambda/t=0.5$ and $U/t=10$ (blue and red indicate poles and zeros of the Green function respectively) along with spinon dispersion multiplied by 10 (black).  Right panel:  momentum-summed spectral weight $A_\uparrow(\omega)=\sum_{\bk} \Tr A_\uparrow(\bk,\omega)/N_s$ with $N_s$ number of unit cells. b) Zoom on the $\bK$ point for different $\lambda$. Black lines show the spinon dispersion. The Chern number  $N_\uparrow$ for spin up associated to the zeros is given in each plot title.}
		\label{fig:spectral_dos_Dirac_cone}
	\end{figure}

	
	 The left panel of Fig.~\ref{fig:spectral_dos_Dirac_cone} a) shows $|\det G_\sigma(\bk,\omega+i0^+)|$ for a large interaction and a non-zero spin orbit coupling parameter $\lambda$, indicating the poles (blue) and zeros (red) of the interacting single-particle Green function, along with the spinons (black).
	We notice that the zeros reside in the middle of the Hubbard gap, the two different branches are gapped and their dispersion resembles that of the spinons.



	
	The Kane-Mele Hubbard model exhibits a topological transition as $\lambda$ is tuned through $\lambda=0$. The two phases may be distinguished via the spin-dependent topological invariant~\cite{ishikawa_magnetic_1986,ishikawa_microscopic_1987,haldane_berry_2004,gurarie_single-particle_2011,essin_bulk-boundary_2011,wang_simplified_2012,blason_unified_2023}:
	\begin{equation}
		\label{spin_Chern_number}
		N_\sigma=\frac{\varepsilon_{\mu\nu\rho}}{24\pi^2}\int_{\omega \bk} \Tr\left[G_\sigma\partial_\mu G^{-1}_\sigma G_\sigma\partial_\nu G^{-1}_\sigma G_\sigma \partial_\rho G^{-1}_\sigma\right],
	\end{equation}
	where  $\int_{\omega\bk}\equiv\int d\omega\int_{\rm BZ}d^2\bk$ and $G$ is the interacting Green function. In the non-interacting limit  $N_\sigma$ corresponds to $ C_\sigma=\sum_{n}\int_{\rm BZ}d^2\bk \theta(-\epsilon_{\bk n\sigma})\Omega_{n\sigma}(\bk)/(2\pi)$ with $\Omega(\bk)=i\braket{\partial_{k_x} u_{\bk n}}{\partial_{k_y} u_{\bk n}}+h.c.$ the Berry curvature in the non-interacting limit and $u_{\bk n}$ being the eigenstates.  In the interacting case $N_\sigma$ can be different from the spin Hall conductance~\cite{slagle_2015}, as recently outlined in Refs.~\cite{blason_unified_2023,gavensky_connecting_2023,setty_electronic_2023,zhao_failure_2023}.  In the Mott insulating phase $N_\sigma$ is determined by the Green function zeros. Interestingly, applying the non-interacting electron formulae to the mean field spinon Hamiltonian gives the same result as Eq.~\ref{spin_Chern_number}, indicating a connection between the topology of zeros and spinons that will be explored more fully below.    
	
	Fig.~\ref{fig:spectral_dos_Dirac_cone}b) presents the evolution of the spectrum of zeros and spinons across the topological transition  at $\lambda=0$. 
	At $\lambda/t=-0.5$ (left panel)  we observe that both the locus of Green function zeros and the spinon dispersion are gapped. At the transition point $\lambda=0$ we observe gapless (Dirac) behavior near $\bK$ and $\bK'$ in the spinon spectrum and in the  locus of Green function zeros. The gapless behavior is protected by the combination of $C_{3z}$ and $C_{2z}\mathcal{T}$ symmetries. For $\lambda>0$ the gaps in the spinon dispersion and zero locus re-open and the sign of the Chern number changes. More explicitly, we found that for each of the two spin blocks both spinons and zeros are described by an effective Hamiltonian which around $\bK,\bK'$ ($\tau=\pm)$ reads: 
		\begin{equation}
			\mathcal H_\tau(\bk) = v( \tau k_x\sigma^x + k_y\sigma^y) + \tau \Delta \sigma^z + O(k^2),
		\end{equation}
where both $v$ and $\Delta$ depend on the interaction strength $U$ and on whether the spinon or Green function zeros are considered. At the topological transition $\Delta\rightarrow 0$. This behavior is further evidence of a deep connection between Green function zeros and the spinon spectrum in topological Mott insulators.
	
	

	
	\noindent
	{\it Edge zeros and spinons --}\label{subsec:edge_zeros}
	We now consider edge states by generalizing the slave-rotor method to allow for site/bond dependent mean field parameters as detailed in the Supplementary Material ~\cite{Suppmat}. In Fig.~\ref{fig:slab_results} we compare results obtained for cylindrical geometries periodic in the direction parallel to one of the lattice vectors, $\mathbf{a}_1$, and either open or periodic in the other direction (see insets to Fig ~\ref{fig:slab_results}).  The periodic case (along ${\mathbf{a}_2}$) shows gapped spinons as well as gapped zeros. The results for open boundary condition reveal instead gapless modes in both the spinon dispersion and the zeros. As shown in the Supplementary Material (Fig.~\ref{fig:zeros_character}) the gapless modes are localized on the boundaries, mirroring the behavior of the non-interacting case but with one important difference: the direction of the edge zeros is reversed compared to the edge spinons and to the non-interacting case. 
	
	
	\begin{figure}
		\includegraphics[width=\linewidth]{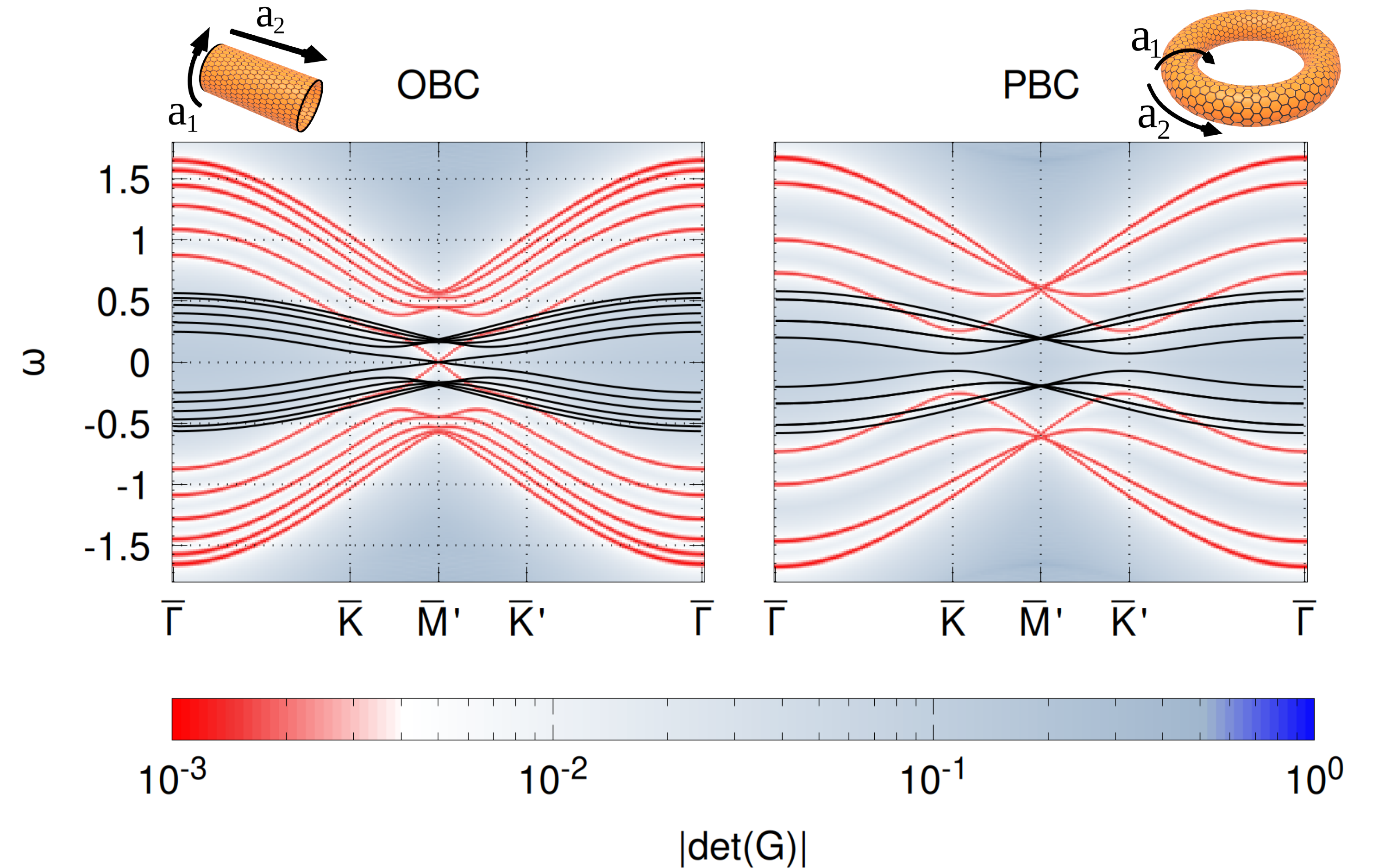}
		\caption{Slave-rotor results for the Kane-Mele-Hubbard model with $\lambda/t = 0.2$ and $U/t=5.5$ on a cylindrical geometry periodic along the crystal $\mathbf{a}_1$ direction and 12 sites along the orthogonal direction ($\mathbf{a}_2$) with either open (OBC, left panel) or periodic (PBC, right panel) boundary conditions. $\overline{ \mathbf{K}}$, $\overline{\mathbf{K}}'$, $\overline{\mathbf{M}}'$ are the projections of the high symmetry Brillouin zone points $\mathbf{K},\mathbf{K}^\prime,\mathbf{M}'$ on the $\mathbf{a}_1$ direction. The color corresponds to the determinant of the Green function and solid black lines indicate the spinon dispersion. In the case with OBC, we find gapless modes with linear dispersion. For each spin block there is one spinon/zero eigenvalue localized at either edge of the cylinder (see Fig.~\ref{fig:zeros_character}).}
		\label{fig:slab_results}
	\end{figure}
	


\begin{figure}[h]
		
		\centering
		\includegraphics[width=1\linewidth]{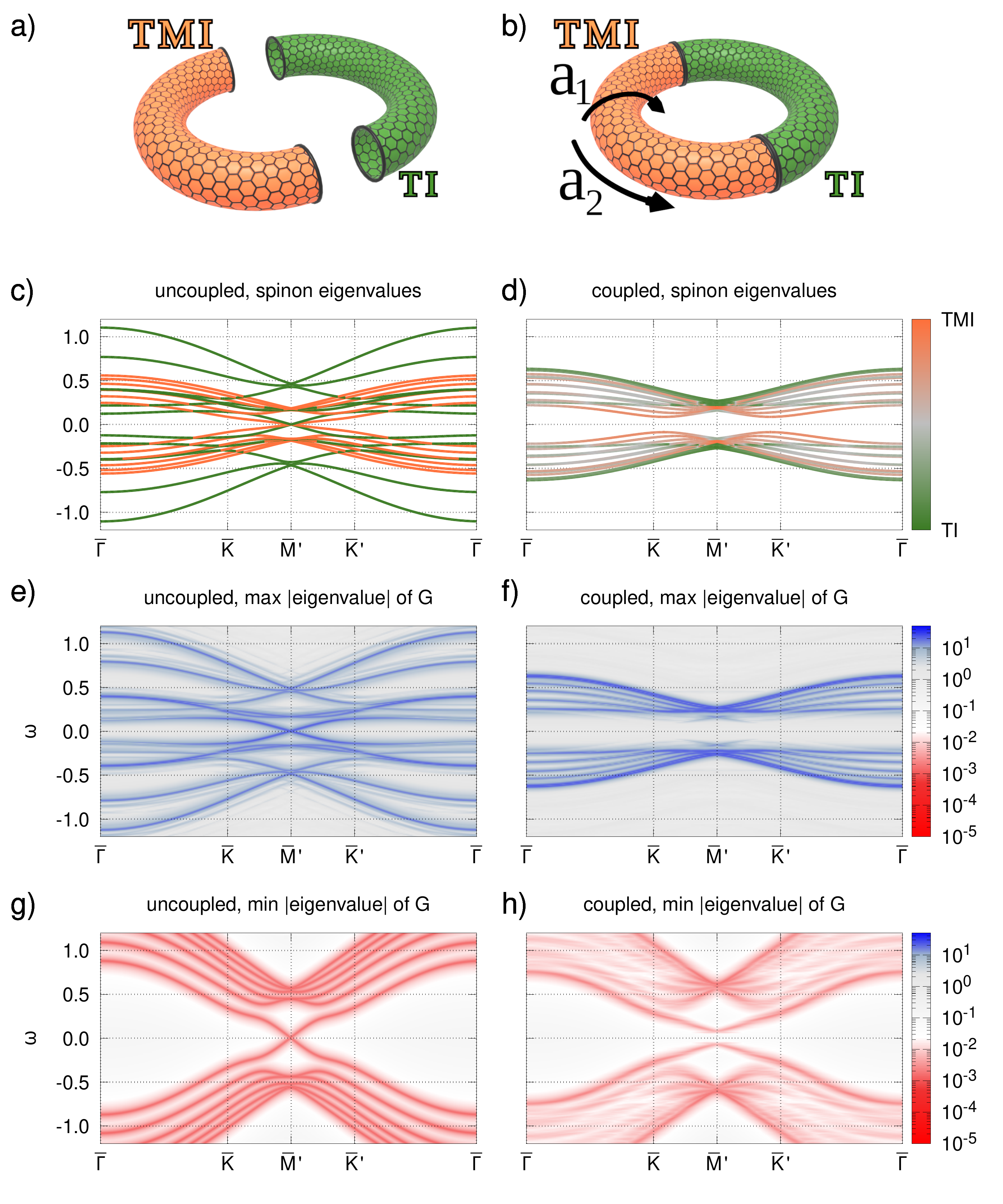}
		
		\caption{a) and b): Sketch of the (un-)coupled Kane-Mele-Hubbard slab geometry. One side, shown in orange, is in the Mott phase ($U/t=5.5$), the other (green) is weakly correlated ($U/t=2.5$). The spin-orbit coupling is set to $\lambda=0.2t$.  Along ${\mathbf{a}_1}$ we consider periodic boundary conditions, i.e. $k_1$ is a good quantum number. Instead, along ${\mathbf{a}_2}$ we consider a finite number of sites, i.e. 24 sites in total. c) and d): spinon eigenvalues for the uncoupled and coupled case. The color indicates on which side of the interface the eigenvalues are localized. e) and f):  maximal absolute value eigenvalue of the slave-rotor Green function for the two sketched cases. g) and h): minimal absolute value eigenvalue. In the disconnected case, c), e) and g), gapless edge spinons, edge poles and edge zeros appear. Upon coupling the two slabs, i.e. going from c), e) and g) to d), f) and h), the edge modes annihilate. }
		\label{fig:interface}
	\end{figure}

	\noindent
	{\it Annihilation of edge zeros and poles -- } 
	Next, we study an interface of a strongly interacting slab which consequently hosts edge zeros and a weakly correlated slab with edge poles. We consider a cylindrical geometry with periodic boundary conditions along ${\mathbf{a}_1}$ and open boundary conditions along ${\mathbf{a}_2}$ such that we have two (identical) interfaces (see Fig.~\ref{fig:interface} a,b). Because of the simultaneous presence of both poles and zeros, the analysis is best made by directly inspecting the eigenvalues of the Green function individually, rather than looking at its determinant. 
	In Fig.~\ref{fig:interface}e-h we show the minimal and maximal eigenvalues of the Green's function, thus resolving zeros and poles respectively. When the two halves of the torus are disconnected (left column), gapless edge poles and edge zeros are evident. Upon putting them in contact with each other (right column), the gapless modes disappear, consistent with a mutual hybridization of the two. This behavior is mirrored for the spinons: in the uncoupled case both sides of the slab host gapless edge spinons (see Fig.~\ref{fig:interface}c), although with different renormalization. When coupled, the gapless edge spinons disappear (Fig.~\ref{fig:interface}d). 
	
	It is possible to understand the annihilation using two different pictures -- either looking at the spinons or the zeros. Considering the spinons, the TMI has a spinon edge mode which at the interface has the opposite direction compared to the edge state of the TI. Consequently, the edge modes cancel each other (as sketched in Fig. \ref{fig:spin-charge-separation}). From the perspective of the zeros the situation is slightly different: the zeros have the same direction at the interface as the TI edge states (see Fig. \ref{fig:zeros_character} and also Ref.\cite{wagner_mott_2023}) and instead of compensating the edge state, they annihilate it.
	
	\noindent
	{\it Spin and charge edge modes separation -- }
 To provide further understanding of the transport properties of the interface between a 2D TI and a 2D Mott TI we consider the  low energy theory of the edge modes of the two systems implied by the slave rotor results. The edge modes of the conventional TI are chiral fermions with chirality defined by the spin and described by the fermionic Hamiltonian $H_{\rm TI} =\sum_\sigma\int dx f^\dagger_{L\sigma}\left(-iv_F\sigma \partial_x\right) f_{L\sigma}$ where $\sigma=\pm$ for spin $\uparrow/\downarrow$. However, the edge modes of the topological Mott Insulator are chiral {\em spinons} described by the Hamiltonian $H_{\rm TMI} =\sum_\sigma\int dx \psi^\dagger_{R\sigma}\left(+iv^*_F\sigma \partial_x\right) \psi_{R\sigma}$ with  Fermi velocity $v^*_F$ determined by the exchange constants in the Mott TI.
 
To understand the coupling at the interface we fractionalize the fermionic edge modes of the conventional TI, writing separately the charge  and spinon parts and we then consider that at the interface the spinon modes are coupled. The time-reversal symmetry and the spin $S^z$ conservation constrain the coupling between the two different edge spinons to $\Delta H= m \int dx \, \sum_\sigma \psi^\dagger_{L \sigma} \psi_{R \sigma} + h.c. $ so that the resulting 1D spinon Hamiltonian, $H_\sigma=\sigma(v_F-v^*_F)k/2+\sigma(v_F+v^*_F)k\mu^z/2+ m\mu^x$ with $\mu$ Pauli matrices in the $L,R$ degrees of freedom,  predicts a  gapped edge spectrum in agreement with the results obtained in our explicit slave rotor calculation.  The  charge modes of the topological insulator, which remain ungapped, provide charge transport confined at the interface between the TI and the TMI, leading to the edge state pattern shown in Fig.~\ref{fig:spin-charge-separation}, where a conventional electronic edge state flows along the free sides of the conventional TI and a conventional spinon edge state flows along the free edges of the Mott TI, but at the interface the spinon modes are gapped, leaving a gapless charge mode to carry the current through the interface. Note that the gap in the spin spectrum means that a physical electron cannot propagate along the Mott TI/TI interface, explaining why the wave packet evolution of a physical electron  presented in Ref.~\cite{wagner_mott_2023} decays as it propagates into the interface. 

The picture just presented is based on a chiral edge mode analysis. In the physical case of unbroken time reversal symmetry with two counterpropagating modes in each system, additional possibilities may occur: in particular residual interactions may lead to gapping of the interface charge modes and charge backscattering at the corner where the free edge of the conventional TI meets the TI/Mott TI interface. Study of this situation is left to a future paper.

	The charge-spin separation taking place at the boundary can be also inferred employing the Ioffe-Larkin composition rules. To this aim we consider the interface of two semi-infinite planes, where the TI edge electrons interact with the edge spinon of the TMI. The longitudinal conductivity along the interface can be readily obtained as $\sigma=\sigma_f+1/(\sigma^{-1}_\psi+\sigma^{-1}_X)$~\cite{ioffe_gapless_1989,aldape_solvable_2022} where $\sigma_f$ is the contribution of the TI edge state, and the contribution from the TMI is comprised by the spinon part $\sigma_\psi$ and the rotor contribution $\sigma_X$. 
	In the topological Mott insulator the rotor degree of freedom $X$ is gapped and, as a result, $\sigma_X=0$. Thus, the conductivity reduces to $\sigma=\sigma_f$ where each spin species contributes with the quantized value $\sigma = e^2/h$. In other words, at the interface only the edge modes of the TI couples to the electromagnetic field and contribute to the charge transport. Finally, assuming to reduce the value of the interaction below the Mott transition, when the slave rotor is condensed $\sigma_X=\infty$, the conductivity becomes $\sigma=\sigma_f+\sigma_\psi$. Both edge modes have finite charge, spin and charge are bound in the electronic quasiparticle and the spin-resolved conductivity vanishes. 
	
	\noindent
	{\it Conclusions --} The results of this paper indicate a deep connection between the dispersion of Green function zeros (a mathematical concept introduced in the context of generalizations of the Luttinger theorem) and the spinon spectrum of a topological Mott insulator. Using slave rotor calculations to study the Kane-Mele Hubbard model we found that in the bulk both the Green function zeros and the spinon spectrum exhibit gaps, that close at the topological transition where the spin-dependent topological index $N_\sigma$ changes sign.
    In the topological phases the Green function zeros and the spinon spectrum exhibit gapless boundary modes. At an interface between a topological Mott insulator and conventional topological insulator the coupling between the Green function zeros/gapless spinons and the fermionic edge state of the conventional topological insulator leads to a novel spin-charge separated state in which the spin degrees of freedom are gapped and charge transport along the boundary is provided by charge $e$ spin zero holon states. 
 
 Our results raise many questions and suggest many directions for future research.  We investigated the large-$U$ regime of a Kane-Mele-Hubbard model focusing on the paramagnetic Mott insulator without discussing the competition with magnetic phases~\cite{ran_direct_2008,hermele_stability_2004,rachel_topological_2010} or charge order/sublattice potential; extending our results to these cases  is of great interest.  Further,  we  restricted ourselves to the particle-hole symmetric case. Breaking particle-hole symmetry by varying an external potential leads to interesting interface structures now under investigation.  The theory of backscattering at the junction where the non-topological and topological edge modes meet is of both theoretical and experimental interest, as is the extension of our considerations to chiral Mott states and fractional Chern states. Going beyond the approximate slave-rotor method used here, and studying a wider range of models is also of great interest; and many-body physics of the coupled edge states may yield further novel interface phases. 
 
	\textit{Note: During the completion of this manuscript we became aware of a related
work \cite{elio}; where there is overlap our results agree.}
 
	\noindent
	{\it Acknowledgments --}
	We acknowledge useful discussions with R.~Raimondi, J.~H.~Pixley,  J.~Cano, L.~Crippa, E.~K\"onig, S. Bollmann, M.~Fabrizio, S.~Sachdev and B.~Trauzettel. N.W. and G.S. are supported by the SFB 1170
	Tocotronics, funded by the Deutsche Forschungsgemeinschaft (DFG, German Research Foundation)  Project-ID 258499086. G.S. acknowledges financial support from
	the DFG through the W\"urzburg-Dresden Cluster of Excellence on Complexity and Topology in Quantum Matter ct.qmat (EXC 2147, project-id 390858490). AJM acknowledges support from the Columbia Materials Science and Engineering Research Center on Precision Assembled Quantum Materials through NSF grant DMR-2011738. The Flatiron Institute is a division of the Simons Foundation. The numerical calculations have been carried out using the Julia language \cite{bezanson_julia_2017}. N.W and G.S gratefully acknowledge the Gauss Centre for Supercomputing e.V. (www.gauss-centre.eu) for funding this project by providing computing time on the GCS Supercomputer SuperMUC-NG at Leibniz Supercomputing Centre (www.lrz.de). This research was supported in part by grants NSF PHY-1748958 and PHY-2309135 to the Kavli Institute for Theoretical Physics (KITP).

	\bibliography{sample}

	\onecolumngrid
	\newpage
	\makeatletter 
	
	\begin{center}
		\textbf{\large Supplementary material for \\ ``\@title ''} \\[10pt]
		
		Niklas Wagner$^1$, Daniele Guerci$^2$, Andrew J. Millis$^{3,2}$ and Giorgio Sangiovanni$^1$   \\
		\textit{$^1$Institut f\"ur Theoretische Physik und Astrophysik and
W\"urzburg-Dresden Cluster of Excellence ct.qmat, Universit\"at W\"urzburg, 97074 W\"urzburg, Germany, Germany} \\
		\textit{$^2$ Center for Computational Quantum Physics, Flatiron Institute, 162 5th Avenue, NY 10010, USA}\\
		\textit{$^3$ Department of Physics, Columbia University, New York, NY 10027, USA}

	\end{center}
	\vspace{20pt}
	
	\setcounter{figure}{0}
	\setcounter{section}{0}
	\setcounter{equation}{0}
	
	\renewcommand{\thefigure}{S\@arabic\c@figure}
	\makeatother
	
	\onecolumngrid
	\appendix
	
	These supplementary materials contain the details of analytic calculations as well as additional numerical details supporting the results presented in the main text.
	
	\section{Conventions}
	\label{app:conventions} 
	
	We choose a system of coordinates in which the scalar product of the $\pm {\bm K}$ vectors with the nearest neighbor vectors ${\bm u}_{j=1,2,3}$ and ${\bm \gamma}_{j=1,2,3}$ are given by
	\begin{equation}
		\label{convention}
		\begin{array}{c||c|c|c|c|c|c} 
			{\bm r} & {\bm \gamma}_1 &  {\bm \gamma}_2 &  {\bm \gamma}_3 &  {\bm u }_1 & {\bm u}_2 &  {\bm u}_3 \\\hline\hline
			e^{i\bK \cdot {\bm r}} & \omega^* & \omega^* & \omega^* & 1 & \omega & \omega^* \\\hline
			e^{i\bK' \cdot {\bm r}} & \omega & \omega & \omega & 1 & \omega^* &  \omega 
		\end{array} \, \, , \quad \omega = e^{2i\pi/3} .
	\end{equation}
	This can be realized, for instance, with ${\bm u}_j =C^{j-1}_{3z}{\bm u}_1$ and ${\bm u}_1=a(1,0)/\sqrt{3}$, ${\bm a}_1 = a(\sqrt{3},1)/2$, ${\bm a}_2 = a(-\sqrt{3},1)/2$, $\bm K= -\bm K' = (4\pi/3a)(0,1)$ and ${\bm \gamma}_1={\bm a}_1+{\bm a}_2$, ${\bm \gamma}_j=C^{j-1}_{3z}{\bm\gamma}_1$. We also introduce the Fourier transform of the hopping process: 
	\begin{equation}
		\label{tunneling_Fourier}
		\begin{split}
			&\epsilon_{\bk}=2\lambda\sum^3_{j=1} \cos{\bm\gamma}_j\cdot\bk,\\
			&\epsilon_{\bk A/B \sigma}=2\lambda\sum^3_{j=1} \cos\left({\bm\gamma}_j\cdot\bk\pm\pi\sigma/2\right),\\
			& V_{\bk}=\sum^{3}_{j=1} e^{i\bk\cdot{\bm u_j}}.
		\end{split}
	\end{equation}

	\section{Details on the slave rotor mapping and on the large-$M$ theory}
	\label{app:slave_rotor}

	In this section we detail the slave rotor mapping, providing the Hamiltonian expressed in terms of the auxiliary representation, the saddle-point equations and the self-consistent expression of the interacting Green function. 
	Employing the slave rotor mapping introduced in the main text Eq.~\eqref{slave_rotor_mapping} the interacting part of the Hamiltonian takes the form:
	\begin{equation}
		\label{rotor_interaction}
		H_{\rm int}=\frac{U}{2}\sum_{s,\br\in s}L^{s\,2}_{\br} .
	\end{equation}
 Hoppings are dressed by a dynamical Peierls phase: 
	\begin{equation}
		\label{t_hopping}
		H_t=-t\sum_{\langle \br\in A,\br'\in B\rangle}e^{-i\left(\theta^A_\br-\theta^B_{\br'}\right)}\psi^{A\dagger}_\br \psi^{B}_{\br'}+h.c.,
	\end{equation}
	and
	\begin{equation}
		\label{SOC_hopping}
		H_{\rm SOC}=\lambda\sum_{s=A,B}\sum_{\langle \br,\br'\rangle\in s} e^{-i(\theta^s_\br-\theta^s_{\br'})}\psi^{s\dagger}_\br e^{i\frac{\pi}{2} \nu_{\br,\br'}\sigma^z}\psi^s_{\br'}.
	\end{equation}
	The mapping is supplemented by the constraint: 
	\begin{equation}
		\label{constraint_total_L}
		L^s_\br-\sum_\sigma \psi^{s\dagger}_{\br\sigma}\psi^s_{\br\sigma}+1=0.
	\end{equation}
	To make progress we turn to the action formalism: 
	\begin{equation}
		\begin{split}
			\mathcal S=&\int^\beta_0 d\tau\sum_{s=a,b}\sum_{\br\in s}\left[-iL^s_{\br}\partial_\tau\theta^s_{\br}+\psi^{s\dagger}_{\br}\left(\partial_\tau -h^s_{\br}\right)\psi^s_{\br}\right.\\
			&\left.+h^s_{\br}\left(L^s_{\br}+1\right)\right] +\frac{U}{2} \sum_{s,r\in s}L^{s\,2}_r+H_0[\theta,\psi],
		\end{split}
	\end{equation}
	where $\tau\in[0,\beta]$ with $\beta=1/T$, $h^s_{\br}(\tau)$ is the field enforcing the constraint in Eq.~\eqref{constraint_total_L}, $H_0[\theta,\psi]=H_\Delta + H_t + H_{\rm SOC} $ is expressed in terms of the spinon $\psi$ and the phase variable $e^{i\theta}$ see Eqs.~\eqref{rotor_interaction},\eqref{t_hopping} and~\eqref{SOC_hopping}, respectively. Integrating out $L^s_{\br}$ we find: 
	\begin{equation}
		\label{lagrangian_angle}
		\begin{split}
			\mathcal L=&\sum_{s,\br\in s}\psi^{s\dagger}_{\br}\left(\partial_\tau -h^s_{\br}\right)\psi^s_{\br}+\frac{(\partial_\tau \theta^s_\br+ih^s_\br )^2}{2U}+h^s_{\br}\\
			&+H_0[\theta,\psi].
		\end{split}
	\end{equation}
	The simplest yet non-trivial treatment of the fluctuations of the theory~\eqref{lagrangian_angle} was developed in Ref.~\cite{florens_quantum_2002,florens_slave-rotor_2004} and consists of introducing the complex boson $X^s_{\br}$: 
	\begin{equation}
		X^s_{\br}=e^{i\theta^s_{\br}},
	\end{equation}
	with the local constraint $|X^{s}_\br(\tau)|^2=1$ which is enforced by the additional Lagrangian multiplier $\lambda^s_{\br}$. Introducing $X$ in the  Lagrangian~\eqref{lagrangian_angle} we find: 
	\begin{equation}
		\label{lagrangian_X}
		\begin{split}
			\mathcal L=&\sum_{s,\br\in s}\psi^{s\dagger}_{\br}\left(\partial_\tau -h^s_{\br}\right)\psi^s_{\br}+\frac{|\partial_\tau X^s_\br|^2-h^{s2}_\br}{2U}+h^s_{\br}\\
			&+\sum_{s,\br\in s}\rho^s_{\br}\left(|X^s_\br|^2-1\right)+\frac{h^s_\br}{2U}\left(X^{s*}_\br \partial_\tau X^s_\br-h.c.\right)\\
			&+H_0[X,\psi].
		\end{split}
	\end{equation}
	Then, the mean field theory can be rigorously derived as the saddle-point of the non-linear sigma model in Eq.~\eqref{lagrangian_X} in the large-$M$ limit~\cite{moshe_quantum_2003}. 
	The limit consists of replacing O(2) with O(2$M$) or, in other words, extending $X^s_{\br}$ to an $M$ components field $X^{s,\alpha}_{\br}$ with $\alpha=1,\cdots,M$ and $\sum^{M}_{\alpha=1}|X^{s,\alpha}_{\br}(\tau)|^2=1$. We also introduce $\alpha=1,\cdots,M$ Fermi fields $\psi^{s,\alpha}_{\br}$ for each spin and  sublattice degree of freedom. Performing the Hubbard-Stratonovich transformation which introduces the bond variables $Q^X$ and $Q^\psi$ we have: 
	\begin{equation}
		\label{lagrangian_X_app}
		\begin{split}
			&\mathcal L_M=\sum_{s,\br\in s}\sum^M_{\alpha=1}\psi^{s,\alpha\dagger}_{\br}\left(\partial_\tau -h^s_{\br}\right)\psi^{s,\alpha}_{\br}-\frac{h^{s2}_\br}{2U}+h^s_{\br}+\sum_{s,\br\in s}\rho^s_{\br}\left(\sum^M_{\alpha=1}|X^{s,\alpha}_\br|^2-M\right)\\
			&+\sum_{s,\br\in s}\sum^M_{\alpha=1}\frac{|\partial_\tau X^{s,\alpha}_\br|^2}{2U}+\frac{h^s_\br}{2U}\left(X^{s,\alpha*}_\br \partial_\tau X^{s,\alpha}_\br-h.c.\right)-\lambda M\sum_{s,\langle \br,\br'\rangle\in s}Q^X_{ss}(\br,\br')Q^\psi_{ss}(\br,\br')\\
			&+\sum_{s,\langle \br,\br'\rangle\in s}\sum^M_{\alpha=1}\left[ Q^\psi_{ss}(\br,\br')\psi^{s,\alpha\dagger}_\br e^{i\frac{\pi}{2} \nu_{\br,\br'}\sigma^z}\psi^{s,\alpha}_{\br'}+ Q^X_{ss}(\br,\br')X^{s,\alpha*}_\br X^{s,\alpha}_{\br'}\right]+tM \sum_{\langle \br\in a,\br'\in b\rangle} Q^X_{ab}(\br,\br')Q^\psi_{ab}(\br,\br')+h.c.\\
			&-t\sum_{\langle \br\in a,\br'\in b\rangle}\sum^M_{\alpha=1}\left[Q^{\psi}_{ab}(\br,\br')\psi^{a,\alpha\dagger}_\br\psi^{b,\alpha}_{\br'}+Q^X_{ab}(\br,\br')X^{a,\alpha*}_\br X^{b,\alpha}_{\br'}+h.c.\right] .
		\end{split}
	\end{equation}
	In the limit $M\to\infty$ the solution is obtained by taking the saddle point of the previous Lagrangian. Assuming that translational invariance and the point group symmetries of the lattice model are preserved, the self-consistency equations associated with $h^s_{\br}$ and $\rho^s_\br$ read:
	\begin{equation}
		\label{constraints}
		\begin{split}
			&\frac{T}{N_s}\sum_{\bk\neq 0,i\nu} G^{ss}_X(\bk,i\nu)+Z^{ss}=1,\\
			&\frac{T}{N_s}\sum_{\sigma,\bk}\sum_{i\omega} G^{ss}_{\sigma,\psi}(\bk,i\omega)=1-\frac{2h^s}{U}-
			\frac{T}{UN_s}\sum_{\bk\neq 0,i\nu}i\nu G^{ss}_X(\bk,i\nu)\left[e^{i\nu 0^+}+e^{-i\nu 0^+}\right],
		\end{split}
	\end{equation}
	with $\omega=(2n+1)\pi T$, $\nu=2n\pi T$, $N_s$ number of unit cells. For all calculations we used a temperature $T/t = 0.01$. $Z^{ss} \equiv Z^s$ is the quasiparticle weight which is determined self-consistently. Note that we do not consider a staggered potential or chemical potential shift and thus the second constraint is automatically fulfilled for $h^s=0$.  In addition, we also have the saddle-point equations for the bond variables $Q^X$:
	\begin{equation}
		\label{QX}
		\begin{split}
			Q^X_{ss}&=\frac{T}{6\lambda N}\sum_{\sigma,\bk}\sum_{i\omega}\epsilon_{\bk s\sigma }G^{ss}_{\sigma,\psi}(\bk,i\omega)e^{i\omega0^+},\\
			Q^X_{ab}&=\frac{T}{3N}\sum_{\sigma,\bk}\sum_{i\omega}V_{\bk}G^{ba}_{\sigma,\psi}(\bk,i\omega) e^{i\omega0^+},
		\end{split}
	\end{equation}
	and $Q^\psi$: 
	\begin{equation}
		\label{Qpsi}
		\begin{split}
			Q^\psi_{ss}&=Z^{ss}+\frac{T}{6\lambda N}\sum_{\bk\neq 0,i\nu}\epsilon_{\bk }G^{ss}_{X}(\bk,i\nu),\\
			Q^\psi_{ab}&=Z^{ab}+\frac{T}{3N}\sum_{\bk\neq 0,i\nu}V_{\bk}G^{ba}_{X}(\bk,i\nu),
		\end{split}
	\end{equation}
        where the $\bk=0$ condensate term is isolated from the finite $\bk$ contribution of the fluctuations. The quantities $\epsilon_{\bk}$, $\epsilon_{\bk s\sigma}$ and $V_{\bk}$ are the Fourier transform of the hopping process and are given in Eq.\eqref{tunneling_Fourier}. In Eq.~\eqref{constraints} and~\eqref{Qpsi} the condensate part~\cite{lee_u1_2005}  is given by $Z^{ss'}=\langle X^{s\dagger}(\bk=0,\tau)X^{s'}(\bk=0,\tau)\rangle/N = \sqrt{Z^s Z^{s'}}$ with $s,s'=A,B$. It is finite in the weakly correlated regime and vanishes in the Mott insulator. The Green functions of the spinon and of the rotor are given by:
	\begin{equation}
		\label{spinon_Green_function}
		G^{-1}_{\sigma,\psi}(\bk,i\omega)=i\omega + h -\begin{bmatrix}
			Q^\psi_{aa} \epsilon_{\bk a\sigma} & -t Q^\psi_{ab} V_\bk\\
			-t Q^\psi_{ba} V^*_\bk & Q^\psi_{bb} \epsilon_{\bk b\sigma}
		\end{bmatrix},
	\end{equation}
	and
	\begin{equation}
		\label{X_boson_Green_function}
		G^{-1}_X(\bk,i\nu)=\frac{\nu^2}{U}- \frac{2i\nu h}{U}+\rho + \begin{bmatrix}
			Q^X_{aa} \epsilon_{\bk} & -t Q^X_{ab} V_\bk\\
			-t Q^X_{ba} V^*_\bk & Q^X_{bb} \epsilon_\bk
		\end{bmatrix}.
	\end{equation}
	For the sake of completeness, we recall that in imaginary time and real space the spinon and rotor Green functions are defined as $G^{ss'}_{\sigma,\psi}(\br-\br',\tau)=-\langle T_\tau \psi^{s}_{\br\sigma}(\tau)\psi^{s'\dagger}_{\br'\sigma} \rangle$ and $G^{ss'}_{X}(\br-\br',\tau)=\langle T_\tau X^{s}_{\br}(\tau)X^{s'\dagger}_{\br'} \rangle$, respectively. In the rotor Green's function $G_X$ we only keep the fluctuation part, the contribution of the condensed fraction has been singled out in Eq.~\ref{constraints} and~\ref{Qpsi}. Finally, within the large-$M$ limit it is straightforward to find that the slave rotor expression of the interacting single-particle Green function reads: 
	\begin{equation}
		\label{single_particle_Green_saddle_point}
		\begin{split}
			G^{ss'}_{\sigma}(\bk ,i\omega) =& Z^{ss'} G^{ss'}_{\sigma,\psi}(\bk ,i\omega) \\
			+ \frac{T}{N} &\sum_{\bq,i\nu} G^{ss'}_{\sigma,\psi}(\bk-\bq,i\omega-i\nu) G^{ss'}_X(\bq,i\nu),
		\end{split}
	\end{equation}
	with $i\nu$ bosonic Matsubara frequency $\nu=2n T$ and $\bq$ momentum transfer between the rotor and the spinon. We emphasize that the first term in Eq.~\eqref{single_particle_Green_saddle_point} is absent in the Mott insulating regime $Z^{ss'}=0$ while the second gives rise to the incoherent Hubbard bands and zeros discussed in the main text.

	\subsection{Numerical solution of the self-consistency equations}
	\label{subsec:numerics}
	
\begin{figure}[h]
		\includegraphics[width=0.5\linewidth]{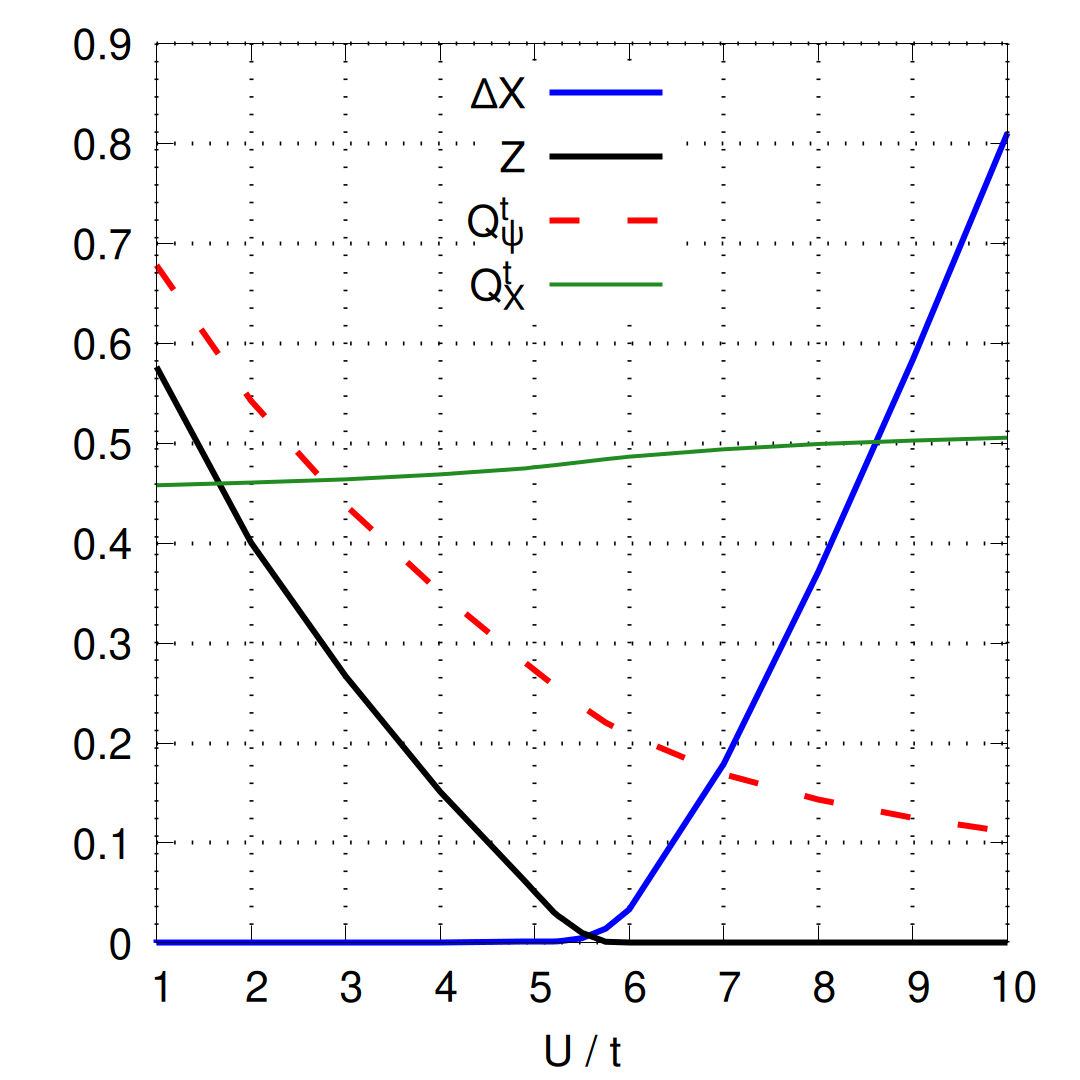}
		\caption{Self-consistent slave-rotor parameters as a function of interaction strength $U$. $\Delta X$ is the gap of the rotors, $Z$ is the quasiparticle weight and $Q_\psi^t$ and $Q_X^t$ are the renormalization of the $t$-hopping term for spinon and rotor respectively. The calculations are done with $\lambda/t = 0.5$. }
		\label{fig:Udep}
	\end{figure}

	
	In this section we provide additional results of the self-consistency equations for periodic and cylindrical boundary conditions. The results for the former are in agreement with earlier studies in Ref.~\cite{rachel_topological_2010,fernandez_lopez_bad_2022}.   
	Fig.~\ref{fig:Udep} shows the evolution of $Q^X$, $Q^\psi$, quasiparticle weight $Z$ and the rotor gap as a function of $U$. The latter is related to the Lagrange multiplier $\rho$ and given by $\Delta X = \rho +\operatorname{min}(\epsilon_X)$ where $\epsilon_X$ is the dispersion of the rotors. We emphasize that the rotor gap grow linearly with $(U -U_c)$ after the Mott transition. 
	Interestingly, the dispersion of the spinon is finite and goes like $t/U$ even for $Z=0$ differently from the results obtained at mean-field level with other slave-particle approaches~\cite{coleman_new_1984,kotliar_new_1986,demedici_orbital-selective_2005,lechermann_rotationally_2007,guerci_unbinding_2017}. This property results from the inclusion of Gaussian fluctuations~\cite{raimondi_lower_1993,ruegg_mathsfz_2-slave-spin_2010,riegler_slave-boson_2020,seufert_breakdown_2021} in the self-consistency equation which are taken into account by the saddle-point equations obtained in the large-$M$ limit. 
 
\begin{figure}[h]
		\includegraphics[width=0.9\linewidth]{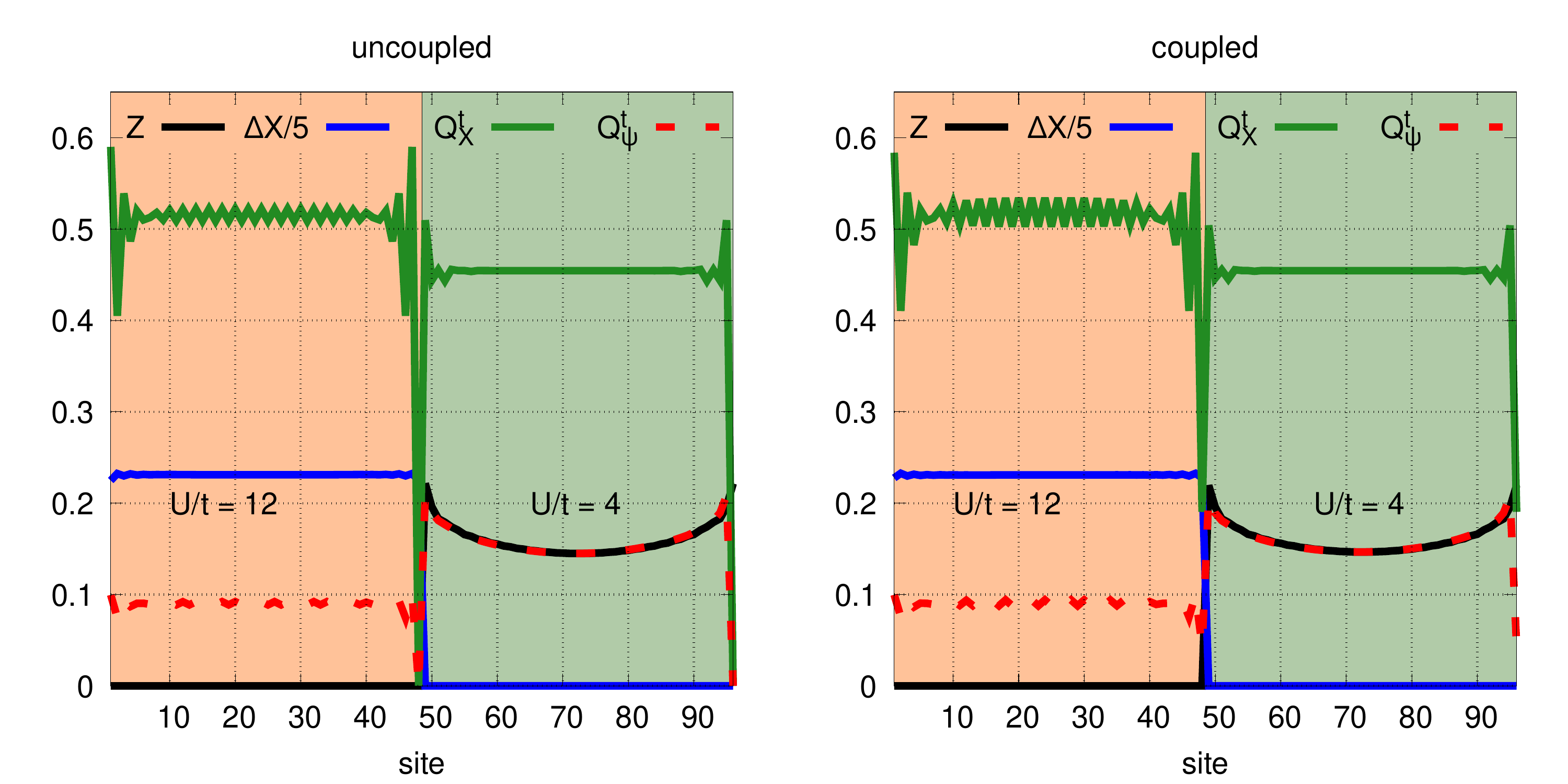}
		\caption{Self-consistent slave-rotor parameters as a function of site for a cylindrical geometry (see Fig.~\ref{fig:interface}) with interface between a TMI ($U/t =12$, orange shade) and a TI ( $U/t = 4$, green shade). $\Delta X$ is the gap of the rotors, $Z$ is the quasiparticle weight and $Q_\psi^t$ and $Q_X^t$ are the renormalization of the $t$-hopping term for spinon and rotor respectively. The calculations are done with $\lambda/t = 0.5$. }
		\label{fig:parameters_interface}
	\end{figure}

\begin{figure}
		
		\centering
		\begin{subfigure}{0.4\textwidth}
			\includegraphics[width=\textwidth]{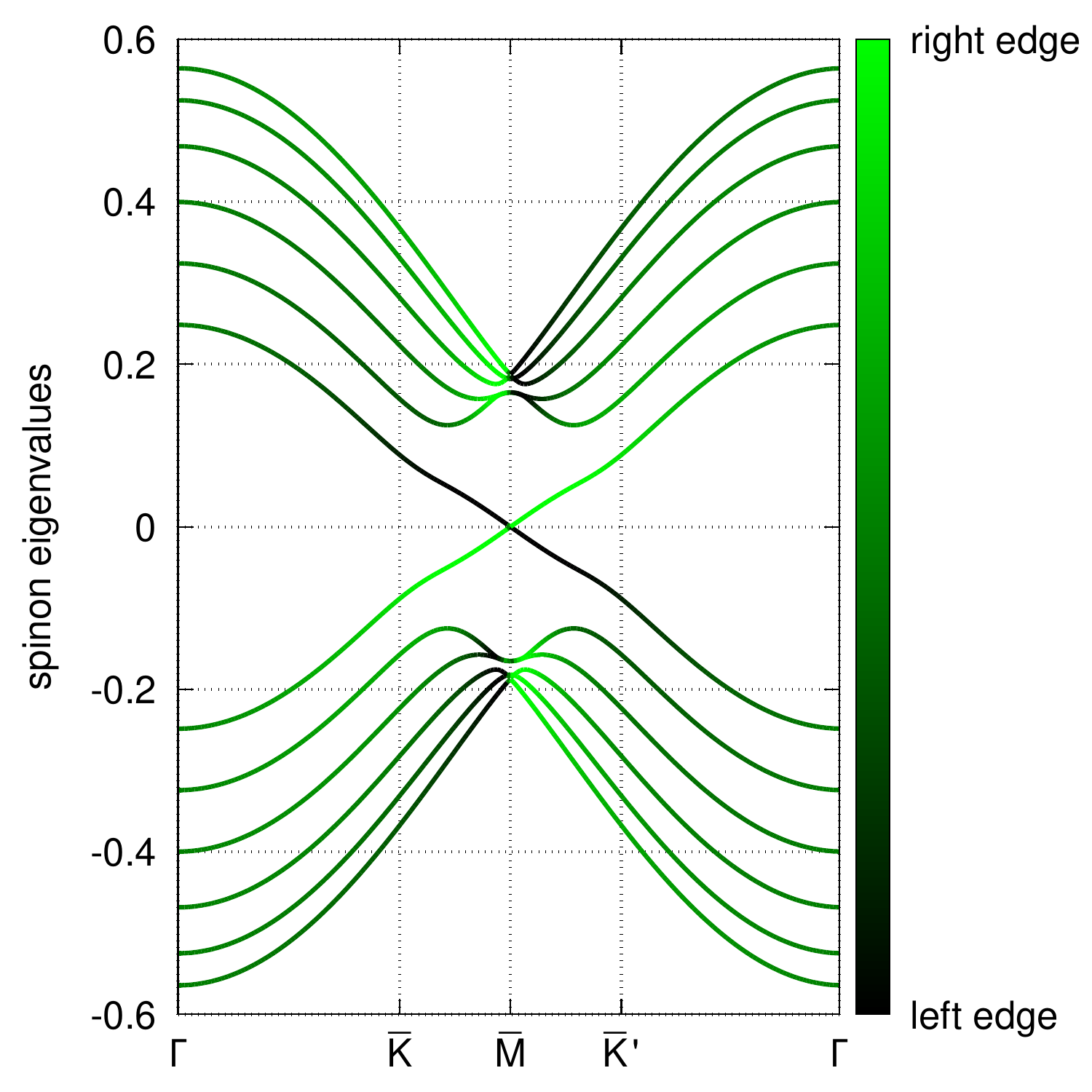}
			\caption{Spinons}
			
		\end{subfigure}
		\begin{subfigure}{0.4\textwidth}
			\includegraphics[width=\textwidth]{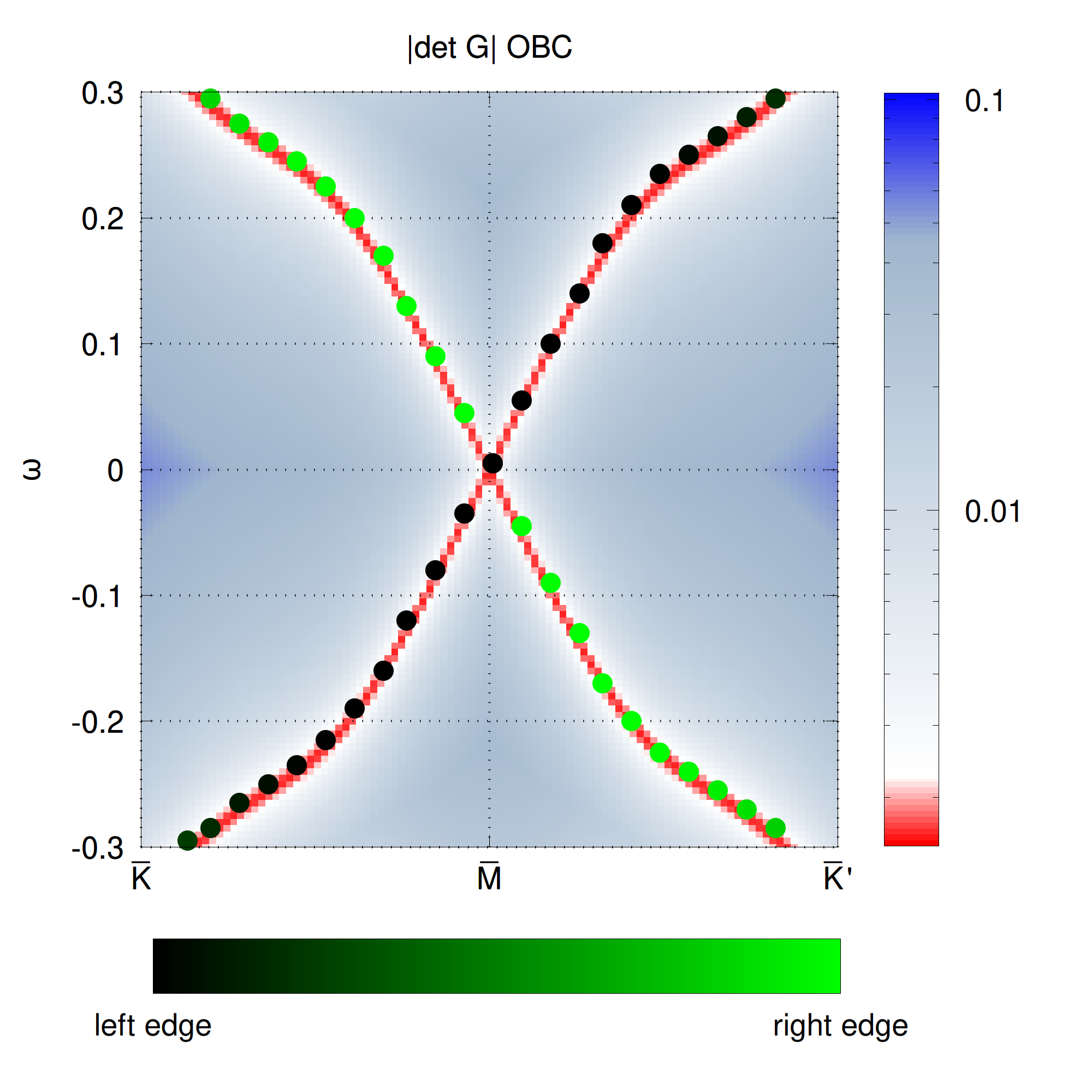}
			\caption{Zeros}
		\end{subfigure}
		
		\caption{
			Localization of edge spinons (a) and edge zeros (b) for the results shown in Fig. \ref{fig:slab_results} . The red-blue color scheme in (b) shows the determinant of the slave-rotor Green function, zoomed-in on the edge state. Black and green dots show the corresponding edge character.}
		\label{fig:zeros_character}
		
	\end{figure}

	Turning to cylindrical boundary conditions, Figure \ref{fig:parameters_interface} shows the site-dependent values of the self-consistent parameters for a slab geometry with interface between a TMI and a TI. As shown there the renormalization parameters vary slightly within the bulk but are mostly unaffected by the interface -- apart from those directly at the interface. Note that the results shown in Fig.\ref{fig:interface} are for a smaller slab, hence the spinon dispersion is more strongly influenced by coupling the two sides.
	Finally, Fig. \ref{fig:zeros_character} shows the localization of edge spinons and egde zeros, revealing the opposite direction of the two edge modes. 
	
\end{document}